\documentclass[apj,numberedappendix]{emulateapj}
\usepackage{apjfonts,graphicx,url,amssymb,amsmath,longtable,rotating,color,units,epsfig}
\newcommand{\be}{\begin{eqnarray}}
\newcommand{\ee}{\end{eqnarray}}

\newcommand{\lp}{\left(}
\newcommand{\rp}{\right)}
\newcommand{\lb}{\left[}
\newcommand{\rb}{\right]}

\newcommand{\msun}{\,M_\odot}
\newcommand{\rsun}{\,R_\odot}

\newcommand{\s}{\,{\rm s}}

\newcommand{\M}{M_1}
\newcommand{\R}{R_1}
\newcommand{\E}{E_{51}}
\newcommand{\K}{\kappa_{0.2}}

\begin{document}

\slugcomment{Submitted for publication in The Astrophysical Journal}

\shorttitle{What Can We Learn from the Rising Lightcurves of SNe?}
\shortauthors{Piro, A. L. \& Nakar, E.}

\normalsize


\title{What Can We Learn from the Rising Lightcurves of Radioactively-Powered Supernovae?}

\author{Anthony L. Piro\altaffilmark{1} and Ehud Nakar\altaffilmark{2}}

\altaffiltext{1}{Theoretical Astrophysics, California Institute of Technology, 1200 E California Blvd., M/C 350-17, Pasadena, CA 91125; piro@caltech.edu}

\altaffiltext{2}{Raymond and Beverly Sackler School of Physics and Astronomy, Tel Aviv University, Tel Aviv 69978, Israel}


\begin{abstract}
The lightcurve of the explosion of a star with a radius $\lesssim
10-100 R_\odot$ is powered mostly by radioactive decay.
Observationally such events are dominated by hydrogen
deficient progenitors and classified as Type I supernovae (SNe I), i.e.,
white dwarf thermonuclear explosions (Type Ia), and core collapses
of hydrogen-stripped massive stars (Type Ib/c). Current transient
surveys are finding SNe I in increasing numbers and at earlier
times, allowing their early emission to be studied in unprecedented
detail. Motivated by these developments, we summarize the physics
that produces their rising lightcurves and discuss ways in which
observations can be utilized to study these exploding stars. The
early radioactive-powered lightcurves probe the shallowest deposits
of $^{56}$Ni. If the amount of $^{56}$Ni mixing in the outermost
layers of the star can be deduced, then it places important
constraints on the progenitor and properties of the explosive
burning. In practice, we find that it is difficult to determine the
level of mixing because it is hard to disentangle whether the
explosion occurred recently and one is seeing radioactive heating
near the surface or whether the explosion began in the past and the
radioactive heating is deeper in the ejecta. In the latter case
there is a ``dark phase'' between the moment of explosion and the
first observed light emitted once the shallowest layers of
$^{56}$Ni are exposed. Because of this, simply extrapolating a
lightcurve from radioactive heating back in time is not a reliable
method for estimating the explosion time. The best solution is to
directly identify the moment of explosion, either through observing
shock breakout (in X-ray/UV) or the cooling of the shock-heated
surface (in UV/optical), so that the depth being probed by the
rising lightcurve is known. However, since this is typically not
available, we identify and discuss a number of other diagnostics
that are helpful for deciphering how recently an explosion occurred.
As an example, we apply these arguments to the recent \mbox{SN Ic
PTF 10vgv.} We demonstrate that just a single measurement
 of the photospheric velocity and
temperature during the rise places interesting constraints
on its explosion time, radius, and level of $^{56}$Ni mixing.

\end{abstract}

\keywords{hydrodynamics ---
    shock waves ---
    supernovae: general}


\section{Introduction}
\label{sec:introduction}

A typical supernova (SN) lightcurve is powered by a combination of
two sources: (1) the energy deposited by the SN shock and
(2) the radioactive decay of $^{56}$Ni that was synthesized during
the explosion\footnote{Here we ignore more exotic energy
sources such as interaction of ejecta with circum-SN matter
(such as in Type IIn SNe) and spindown of a rapidly rotating
magnetar \citep{KasenBildsten2010,Woosley2010}.}. The first light from shock breakout and
the following early emission (for the first minutes to days) is
dominated by the shock-deposited energy. The very late
emission, after the radiation can efficiently diffuse through the
entire ejecta ($\gtrsim 100$ d), is dominated by radioactive decay.
The relative influence of these power sources during the time in
between depends on two main factors: (1) the progenitor radius
$R_*$ and (2) the amount of $^{56}$Ni. The reason for this is that
the bolometric luminosity from shock heating increases linearly with
the progenitor radius (as we discuss in \S
\ref{sec:shockheating}),
while the peak of
the radioactively-powered luminosity is roughly linear with the
total mass of $^{56}$Ni \citep{Arnett79}.

Core-collapse SNe typically synthesize $\sim0.01-0.1M_\odot$ of
$^{56}$Ni, and the progenitor radii range between
$\sim1-1000R_\odot$. Therefore, in more extended progenitors the
main SN event is dominated by shock heating, while in more compact
ones it is dominated by radioactive power. The most common SNe II-P
are explosions of red supergiants with $R_* \sim 500R_\odot$. They
exhibit an extended plateau phase for about a hundred days, which is
powered by the cooling of shock-heated material. The progenitors of
the rare 1987A-like SNe II are blue supergiants with $R_* \sim
50R_\odot$ \citep{Woosley88,Kleiseretal11}, and their lightcurves
are dominated by radioactive decay starting a week after the
explosion. Similarly, SNe IIb with compact progenitors ($R_*\lesssim10\ R_\odot$) like
SN 2011dh \citep{Arcavietal2011} show evidence
of both shock-heated material and a separate radioactive peak.
The progenitors of SNe Ib/c are massive stars that were
stripped of their hydrogen envelope and are also fairly
compact with $R_* \lesssim 10R_\odot$. Indeed SNe Ib/c are
dominated by radioactive power starting a few days (or earlier) after the
explosion. Finally, for the compact
white dwarf progenitors of SNe Ia with $R_*\lesssim0.01R_\odot$, the
shock-heating lightcurves are so dim that they have never been
observed, and our knowledge of these events is only possible due to
the synthesis of $\sim0.5M_\odot$ of $^{56}$Ni.



Since the physics that governs the lightcurves of most SNe
with $R_* \lesssim 50R_\odot$ has many similarities, a single
theoretical framework should roughly describe their main qualitative
features. Their emission during the first few days is especially exciting because it probes the shallowest
layers of the progenitor. This can teach us about the exploding
star's radius, and constrain the surface composition and
velocity/density gradients that reflect details of the explosive
burning. The shock-heating contribution in SNe I has been
well-studied in the literature, both with semi-analytic models
\citep{NakarSari2010,RabinakWaxman2011} and with detailed radiative
transfer simulations \citep{Dessartetal2011}. The rising lightcurve
from radioactive heating has been explored for the cases of SNe Ia
\citep{Piro2012} and to study the impact of $^{56}$Ni mixing in SNe Ib/c
\citep{Dessartetal2012}.


Concurrent with these theoretical studies, transient
surveys like with the Katzman Automatic Imaging Telescope \citep[KAIT;][]{Filippenkoetal2001}, and by
the Palomar Transient Factory
\citep[PTF;][]{Rauetal2009,Lawetal2009} and the Panoramic Survey
Telescope and Rapid Response System
\citep[Pan-STARRS;][]{Kaiseretal2002} are finding increasing numbers
of these events, especially at early times. Best known among these is SN 2011fe,
the closest SNe Ia in the last 25 years
\citep{Nugentetal2011,Bloometal2012}. Other SNe Ia reported within the last year
with  early data include SN 2009ig
\citep{Foleyetal2012}, SN 2010jn \citep{Hachingeretal2012}, and SN
2012cg \citep{Silvermanetal2012}. SNe Ib/c have also been
increasingly well-studied in the optical at early times, as
summarized by \citet{Droutetal2011}. Other particular recent events include
SN 2008D \citep{Soderbergetal2008,Modjazetal2009} and
PTF 12gzk \citep{BenAmietal2012}.
Motivated by these exciting
developments, we ask: {\em what can and what
cannot be learned from the early optical lightcurve of
radioactively-powered SNe?} Being more abundant, we focus
on parameters that are typical to SNe I, although most of our
conclusions can also be applied to subclasses of SNe II that are dominated by radioactive heating.

In our study we first consider instances where only a photometric
lightcurve is available, as is often the case at early times. One of
the main issues is how much we can infer about $^{56}$Ni
mixing in the outermost layers of the star with this limited
information. Our main conclusion is that it is hard to determine the
mixing of $^{56}$Ni based on the lightcurve alone without additional
information. The reason is that the lightcurve can provide an estimate for the {\em total} mass
of $^{56}$Ni that is exposed to the observer at any given time, but
it cannot provide a good handle on the {\em fractional} mass of
$^{56}$Ni with respect to the total exposed mass (which can be
determined if the time of explosion is well constrained, e.g., by
detection of the shock breakout or the cooling envelope phase).
Putting it differently, there is a degeneracy in the lightcurve
between a recent explosion with a high fraction of $^{56}$Ni in the
outermost layers and an older explosion where $^{56}$Ni resides
only in deep material. In the latter case the SN has a ``dark phase''
that can persist for up to a few days after the explosion. During
this time no $^{56}$Ni is exposed and its shock-heated cooling
emission is often too faint for detection even when deep
observations exist. For this reason, a simple extrapolation of the
lightcurve to early times cannot reliably constrain the time of
explosion.

We then consider what additional information can be extracted by
using limited spectral data that provide the photosphere temperature
and velocity. We show that if detailed color evolution, or if one or
even better two spectra (possibly of low signal to noise) are
available during the rise, then the degeneracy  between explosion
time and $^{56}$Ni depth can be at least partially alleviated. Key
among our results is \mbox{equation (\ref{eq:tmin}),} which allows one
to estimate a lower limit on the time of explosion using merely a
single simultaneous measurement of the bolometric luminosity,
temperature, and photospheric velocity. The methods we discuss are
complementary to detailed spectral studies that probe element mixing
based on their absorption and emission features in the rare cases
where early high signal to noise spectra are available
\citep[e.g.,][]{Saueretal2006,Parrentetal12}.

In the following we begin in \S \ref{sec:summary} by discussing
each of the ingredients that shape 
the electromagnetic emission starting less than an hour and up to
weeks after explosion. This discussion provides a useful general
guide for interpreting early observations of radioactively-dominated
SNe. In \S \ref{sec:10vgv} we investigate a specific SN Ic in some
detail (PTF 10vgv) as a test case for applying these arguments and
techniques. We conclude in \S \ref{sec:conclusion} with a discussion
of our results and a summary of important conclusions that should
help facilitate better constraints from future SNe I observations.

\section{Early Emission from Type I Supernovae}
\label{sec:summary}

\begin{figure}
\epsscale{1.0} \plotone{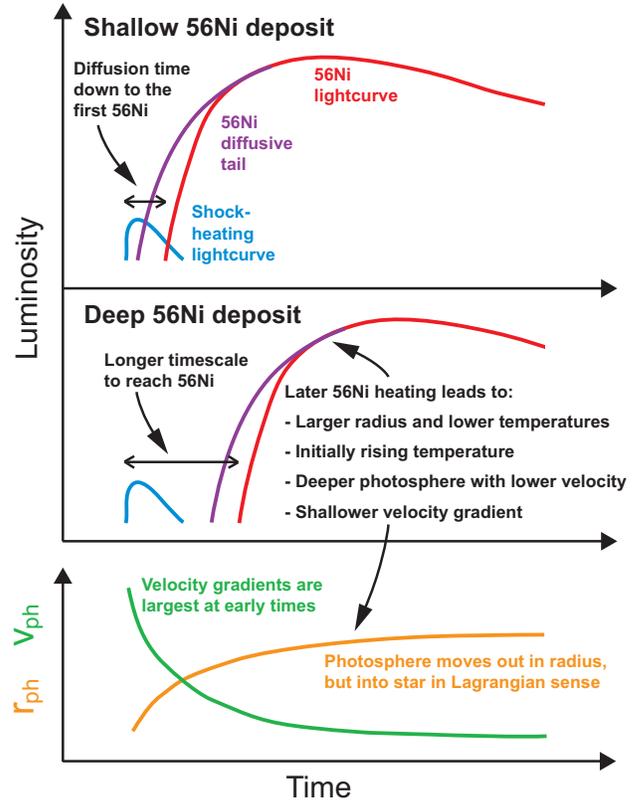} \caption{Schematic diagram showing the early lightcurves for SNe I and how
they relate to the photospheric radius and velocity. The top and
middle panels demonstrate how the relative positions of the
shock-heating lightcurve (blue curves), diffusive tail lightcurve
(purple curves), and the $^{56}$Ni lightcurve (red curves) can
differ depending on the depth of the $^{56}$Ni. The total observed
lightcurve is the sum of these three components. Note that when the
$^{56}$Ni is deposited deeply (in the middle panel) and the
shock-heating lightcurve (blue curve) is below the detection limits,
there can be a significant dark phase between the time of explosion
and the moment of first detection.
In the bottom panel we show how the photospheric radius (orange curve) and velocity (green curve) evolve with time. Depending on the position of the $^{56}$Ni lightcurve, different photospheric radii, velocities, and velocity gradients will be present when the rising lightcurve is observed. These provide clues about the depth of the $^{56}$Ni and the time of explosion, as summarized in the middle panel and discussed in the text.}
\label{fig:diagram}
\epsscale{1.0}
\end{figure}

When a hydrogen-poor star explodes as a SN, a few important events
occur in the moments before and after the first optical emission is
seen. In this section we summarize the main properties of each of
these events, their observational consequences, and how detections
of some or all of these events can be used to put constraints on the
properties of the exploding star. To guide the discussion we will be
referring to the diagram in Figure \ref{fig:diagram}, which shows
the time-dependent luminosity components, photospheric radius, and
velocity.

\subsection{Shock Breakout}

Just prior to emission, a shock is traveling through the envelope.
This heats and accelerates the material, unbinding it from
the star. A radiation-dominated shock accelerates in the
decreasing density of the outer edge of the star with a shock
velocity that scales with the density as  $v_s\propto \rho^{-\beta}$
where $\beta \approx 0.19$ \citep{Sakurai1960}. The shock continues
to shallower regions until the optical depth falls to $\tau\approx
c/v_s$, where c is the light speed. At this point the photons are
no longer trapped; they stream away and the shock dies. This is what
is typically referred to as the ``shock breakout'' UV/X-ray flash,
and it has been frequently studied because of its strong dependence
on the radius (which determines both the energy budget of the shock
and the timescale of the emission), allowing the progenitor star to
be studied from its detection
\citep{Colgate1974,Falk1978,KleinChevalier1978}.

Recently it was realized that the radiation behind the shock falls
out of thermal equilibrium for small radii hydrogen-stripped massive
stars with high shock velocities, and that the breakout will be in
X-rays \citep{Katzetal10,NakarSari2010}. In SNe Ia, the shock
achieves relativistic velocity, and its breakout emission is in the
MeV range \citep{NakarSari12}. In any case, although shock breakout
is the first indication of the explosion, its impact for SNe I in
the optical/UV bands is negligible. For this reason it is not
discussed further in this paper and not  plotted in \mbox{Figure
\ref{fig:diagram}.}

\subsection{Shock-Heated Cooling Lightcurve}
\label{sec:shockheating}

Immediately after shock breakout, the observed radiation is out of
thermal equilibrium until roughly the time when the gas doubles its
radius at $t\approx R_*/v_f$ \citep{NakarSari2010}, where
$v_f\approx 2v_s$ is the final velocity of the material
\citep{MatznerMcKee1999}, which is within minutes or less. The
observed photons then gain thermal equilibrium and the expansion
enters its spherical homologous phase
\citep{Chevalier92,Piroetal2010,NakarSari2010,RabinakWaxman2011}. As
the material that has been heated by the shock expands, a thermal
diffusion wave begins backing its way through the ejecta. Above the
depth of the diffusion wave, material cools via photon diffusion.
Below this depth, material evolves adiabatically. For each fluid
element, there is then a competition between adiabatic cooling and
diffusive cooling that controls the energy density of that material
at the moment its photons begin streaming out of the star. This
determines the observational signature of shock-heated cooling,
leading to a bolometric luminosity
\citep{NakarSari2010}\footnote{The luminosity pre-factor of
\cite{NakarSari2010} is divided here by a factor of 2.5 following
the numerical result of \cite{Katzetal12}}
\begin{equation}
    L \approx 2 \times 10^{41} \frac{\E^{0.91}\R}{\K^{0.82}\M^{0.73}}
    \left(\frac{t}{1 {\rm~hr}}\right)^{-0.35} {\rm erg\ s^{-1}},
    \label{eq:L_coolingPhase}
\end{equation}
and an observed (color) temperature of
\begin{equation}
    T_c \approx 4 ~\frac{\E^{0.11}\R^{0.38}}{\K^{0.23}\M^{0.11}}
    \left(\frac{t}{1 {\rm~hr}}\right)^{-0.61} {\rm eV},
 \label{eq:T_coolingPhase}
\end{equation}
where $E=E_{51}10^{51}\ {\rm erg}$ is the explosion energy, $M=M_1M_\odot$
is the ejecta mass, $R_1=R_*/R_\odot$, and $\kappa$ is the opacity with
$\kappa_{0.2}=\kappa/0.2\ {\rm cm^2\ g^{-1}}$ being the canonical
value during the cooling phase for fully
ionized hydrogen-free gas. For all the scalings in this paper we
assume a polytropic index of $n=3$, as is relevant for compact,
radiative stars forming SNe Ib/c, or relativistic, degenerate WDs exploding as SNe
Ia. The key result is that the luminosity during the shock-heated
cooling is directly proportional to the progenitor radius.

As long as the opacity is dominated by scattering the
observed temperature is determined at the thermalization
depth, below the photosphere, where the optical depth is $\tau>1$. The thermalization optical depth, $\tau_c$, and the
absorption optical depth, $\tau_{\rm abs}$, satisfy at the
thermalization depth  $\tau_c \tau_{\rm abs}\approx 1$.
\mbox{Equation (\ref{eq:T_coolingPhase})} assumes that $\tau_{\rm
abs}$ is dominated by free-free and minor correction are expected
when bound-free absorption dominates \citep[for a detailed
discussion, see][]{NakarSari12}.

In the optical/UV, the luminosity is rising as long it is
in the Rayleigh-Jeans tail, resulting in (see \mbox{Appendix
\ref{sec:rayleighjeanstail}})
\begin{equation}\label{eq:F_opt_coolingPhase}
    L_{\rm opt/UV} \propto t^{1.5}.
\end{equation}
When optical/UV luminosity is observed to rise more steeply than this \citep[as in PTF 12gzk;][]{BenAmietal2012},
it indicates that the rise from shock-heated cooling is not being seen, and the time of explosion is
not confidently constrained.
This phase continues for several hours in a core-collapse SN until
$T_c$ crosses the UV (at which point the UV flux starts to drop) and
reaches the optical band. This is also the point that recombination
starts in cases that radioactive decay does not play an important
role. In SNe Ia this phase is terminated after about an hour when
the diffusion wave reaches material where the shock was
not radiation dominated and the cooling envelope emission drops
significantly \citep{Rabinaketal2011}.

Once $T_c$ drops sufficiently in a core-collapse SN, a
recombination wave begins backing its way in from the surface. The temperature drop
becomes more gradual, settling at \mbox{$\approx5000-8000\ {\rm K}$} and the
luminosity drop stops or even gently rises \citep[as discussed in][also see Goldfriend, Sari
\& Nakar, in preparation]{Dessartetal2011}. As far as optical photometry
is considered, both the temperature and the luminosity are roughly
constant and the SN enters a ``plateau phase.'' This plateau is similar
to the one observed in SN II-P \citep{Popov1993,KasenWoosley2009},
except that for SNe I it is dimmer and
that it starts earlier, both due to the smaller progenitor and
enhanced adiabatic losses. Setting $T_c\approx0.6\ {\rm eV}$ as found for the plateau in the SN Ib/c models of \citet{Dessartetal2011},
equation (\ref{eq:T_coolingPhase}) implies
that the optical plateau starts at
\begin{equation}\label{eq:t_plateau}
    t_p \approx 20
    ~\frac{\E^{0.18}\R^{0.62}}{\K^{0.38}\M^{0.18}} {\rm~hrs},
\end{equation}
and together with equation (\ref{eq:L_coolingPhase}) we find that
\begin{equation}
 L_p \approx 7\times10^{40} ~\frac{\E^{0.85}\R^{0.78}}{\K^{0.69}\M^{0.67}}  {\rm ~erg\s^{-1}},
 \label{eq:L_plateau}
\end{equation}
is roughly the plateau luminosity\footnote{Applying the same line of
arguments to a red supergiant with $R_*=500 \rsun$, $M=15 \msun$, and $\kappa=0.34 ~{\rm cm^2\ g^{-1}}$, results in $t_p
\approx 10\ {\rm days}$ and \mbox{$L_p\approx 10^{42} {\rm
~erg\s^{-1}}$,} compatible with observations of SNe II-P.}.

To conclude, the detection of the shock-heated cooling phase in
core-collapse SNe I is very challenging given its low luminosity and
short duration, but the rewards are also high. Photometry alone of
the rising phase provides tight constraints on the explosion time,
and both the optical rising and subsequent plateau phase provide
constraints on the progenitor radius (if not overcast by radioactive
heating as we discuss next). A convenient method to identify that a
rising optical emission is due to shock-heated cooling emission (and
not radioactively powered emission), is that the optical emission
rises (as predicted by eq. [\ref{eq:F_opt_coolingPhase}]), while the
bolometric luminosity drops. This can be seen, for example, by
having simultaneous UV and optical coverage in the time that $T_c$
sits between these two frequency windows.

\subsection{$^{56}$Ni Shallower than the Diffusion Depth}
\label{sec:shallower}

If there was no radioactive energy input, this would be the end of
the story for the electromagnetic signal from SNe, and most SNe I
would be too dim to have ever been detected. But eventually the
radioactive decay of $^{56}$Ni starts heating the expanding ejecta, and this
is where their lightcurves begin rising in earnest.

When the thermal diffusion wave reaches the shallowest deposits of
$^{56}$Ni, the energy generation from $^{56}$Ni roughly goes
directly into the observed bolometric luminosity (as shown with the
red curves in Figure \ref{fig:diagram}), so that
\be
    L_{56}\approx M_{56}\epsilon,
    \label{eq:luminosity56}
\ee
where $M_{56}$ is the mass of $^{56}$Ni that is exposed by the diffusion wave, and the specific heating rate from $^{56}$Ni is
\be
    \epsilon(t) = \epsilon_{\rm Ni} e^{-t/t_{\rm Ni}} + \epsilon_{\rm Co} ( e^{-t/t_{\rm Co}}- e^{-t/t_{\rm Ni}} ),
    \label{eq:e56}
\ee
where $\epsilon_{\rm Ni}=3.9\times10^{10}\ {\rm erg\ g^{-1}\
s^{-1}}$, $t_{\rm Ni}=8.76\ {\rm days}$, $\epsilon_{\rm
Co}=7.0\times10^9\ {\rm erg\ g^{-1}\ s^{-1}}$, and $t_{\rm
Co}=111.5\ {\rm days}$. It should be noted that in our more detailed analysis in
Appendix \ref{sec:tail}, we show that if $L_{56}$ is the observed luminosity, then
equation (\ref{eq:luminosity56}) is only accurate up to a factor of $\sim2$. Even with
this small uncertainty, the rough level of $^{56}$Ni mixing that can be
deduced from the observations is still robust as we will discuss.

During this phase the depth of the diffusion wave is important
because it tells us which part of the exploding star is being probed
by the observations. The depth in mass is
related to the time after explosion by (see \mbox{Appendix
\ref{sec:diffusiondepth}}) \be
    \Delta M_{\rm diff} \approx 8\times10^{-2}
    \frac{E_{51}^{0.44}}{\kappa_{0.1}^{0.88}M_1^{0.32}}
    \lp \frac{t}{1\ {\rm day}} \rp^{1.76}M_\odot,
    \label{eq:mdiff}
\ee
where $\kappa_{0.1}=\kappa/0.1\ {\rm cm^2\ g^{-1}}$ is the
canonical value we use for the radioactively powered phase, since
the gas is partially ionized. For the estimates in this paper, we
use quantities that roughly correspond to the SNe Ib/c, but the same
general arguments apply to SNe Ia (as well as the rare SNe II that are
radioactively powered) with different prefactors. To see how
much these can vary for other progenitors see \mbox{Appendix
\ref{sec:diffusiondepth}.} Also note that equation (\ref{eq:mdiff}) becomes
less accurate closer to the peak of the lightcurve, when $\Delta M_{\rm diff}$ grows
more slowly with time.

When powered by radioactive heating, the observed luminosity is no longer sensitive to the
progenitor radius as can be seen
by the lack of an explicit dependence on $R_*$ in equation (\ref{eq:mdiff}).
Instead, it provides a direct measurement of the
amount of $^{56}$Ni (via eq. [\ref{eq:luminosity56}]) at the
location of the diffusion wave. If the explosion time is
known, equation (\ref{eq:mdiff}) provides $\Delta M_{\rm diff}$ into which this
$^{56}$Ni is mixed. From this the mass fraction $X_{56}\approx M_{56}/\Delta M_{\rm diff}$ can
be inferred as a function of the depth $\Delta M_{\rm diff}$.
Thus, in principle, detection of the
rise of the $^{56}$Ni lightcurve should provide an estimate of the
distribution of $^{56}$Ni \citep[as attempted for the
SN Ia 2011fe by][]{Piro2012}, which is helpful for understanding
the nature of the explosive burning and the outer structure of the progenitor.


\subsection{$^{56}$Ni Deeper than the Diffusion Depth: ``Diffusive Tail''}
\label{sec:deeper}

In practice, the exercise described above is not as straight forward
as one might think. The reason is that if there is a steep increase
in the $^{56}$Ni abundance, then the assumption that the observed
emission is generated only by the composition at the location of the
diffusion wave may not be valid. Instead a ``diffusive tail''
of the energy released in $^{56}$Ni-rich layers {\em deeper} than
the diffusion depth (shown as purple curves in Figure
\ref{fig:diagram}) may actually dominate over the energy released in
$^{56}$Ni-poor layers {\em shallower} than the diffusion depth
(shown as red curves in Figure \ref{fig:diagram}). Next we
calculate the contribution of this diffusive tail.

Consider a deposit of $^{56}$Ni at some depth $d$ in the star. The
$^{56}$Ni decays to produce gamma-rays, which are absorbed and
create thermal photons. The photons then spread due to
diffusion, creating roughly a Gaussian distribution around the
$^{56}$Ni. At each time $t$, the Gaussian has a width $\sqrt{K}t$,
where $K$ is the diffusion coefficient, which is related to the
diffusion time by $t_{\rm diff}= d^2/K$. For such a
distribution, there is a small, but non-zero, fraction of photons
that escape from the star because they have diffused by a distance
$>d$ from the $^{56}$Ni depth. The fraction of photons that reach
the surface at time $t$ is
\be
    {\rm Escaping\ fraction} \approx {\rm erfc}\sqrt{t_{\rm diff}/2t},
\ee
where ${\rm erfc}$ is the complementary error function.
Therefore, if at some time $t'$ the diffusion wave reaches a layer
of $^{56}$Ni that would be producing a luminosity $L_{56}(t')$ according
to equation (\ref{eq:luminosity56}), this implies
for previous times $t<t'$ that the diffusive tail from this layer
also produces a luminosity. Since $t_{\rm diff}\propto \rho r^2
\propto 1/t$, this diffusive tail luminosity is
\be
    L_{\rm tail}(t<t') = L_{56}(t')
    \frac{\epsilon(t)}{\epsilon(t')}
    \frac{{\rm erfc}( t'/\sqrt{2}t )}{ {\rm erfc}(1/\sqrt{2} )}.
    \label{eq:tail}
\ee
A simpler approximation to this scaling is
\be
    L_{\rm tail}(t<t') \approx L_{56}(t')
    \frac{\epsilon(t)}{\epsilon(t')}
    \frac{t}{t'}\frac{e^{-t'^2/2t^2}}{e^{-1/2}},
    \label{eq:tailapprox}
\ee
which is accurate within $\approx20\%$ for $t'/2<t<t'$ even though $L_{\rm tail}$ changes by a factor of $\approx9$ over this range of times.

\begin{figure}
\epsscale{1.0}
\plotone{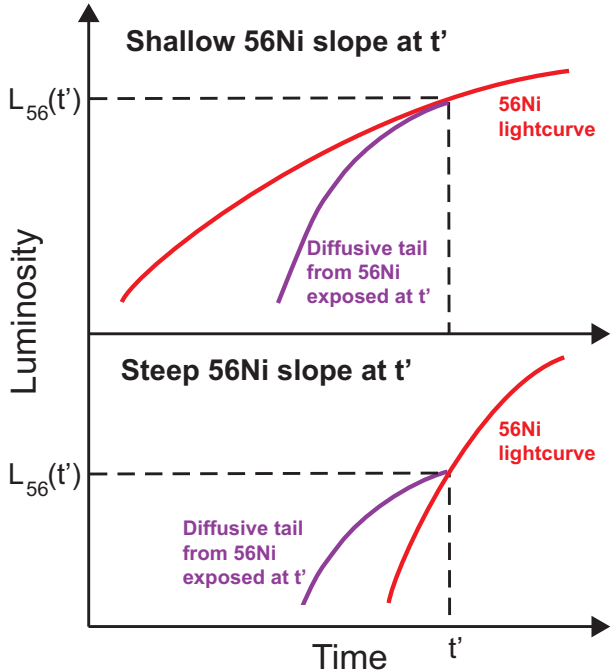}
\caption{Schematic diagram showing how the slope of the $^{56}$Ni distribution determines the relative importance of the diffusive tail. In both the top and bottom panels, a luminosity $L_{56}(t')$ would be produced at time $t'$ from the diffusion wave probing the $^{56}$Ni distribution. In the top panel, the slope of the $^{56}$Ni is shallow, which produces a shallow slope for $L_{56}$ (shown by the red curve). If a diffusive tail luminosity is drawn back from this point (shown by the purple curve), it always falls below the $^{56}$Ni lightcurve. Thus the diffusive tail from this depth is not important. In the bottom panel, the slope of the $^{56}$Ni distribution is steeper. Therefore, the diffusive tail from larger depths overpowers the heating from $^{56}$Ni at shallower depths, and the diffusive tail from this depth impacts the lightcurve. The logarithmic slope of the diffusive tail near the location $t\approx t'$ is roughly $1+(t'/t)^2\approx 2$.}
\label{fig:diagram2}
\epsscale{1.0}
\end{figure}

From this discussion one can see what we meant earlier in this section when
we mentioned ``a steep increase in the abundance of $^{56}$Ni.'' Consider
the two $^{56}$Ni distributions shown in Figure \ref{fig:diagram2}. In
the top panel the $^{56}$Ni has a rather shallow distribution which produces
the $L_{56}(t)$ (shown by the red curve). Therefore if
we consider the luminosity $L_{56}$ at some time $t'$, and then trace back
the diffusive tail implied from that depth using equation (\ref{eq:tail}) (shown
as a purple curve), the diffusive tail always falls below the $L_{56}$ lightcurve.
This is a case where the shallow $^{56}$Ni prevents the diffusive
tail from having a noticeable impact. In the bottom panel the $^{56}$Ni
has a steeper distribution. Now, when the diffusive tail is drawn back
from a point at time $t'$, it exceeds the $L_{56}$ lightcurve. In this latter
case the diffusive tail will dominate the observed rise.

Therefore to evaluate whether the diffusive tail is important at a given depth
the key quantity to consider is the slope of $L_{\rm tail}$. Using equation (\ref{eq:tailapprox}),
the logarithmic derivative is
\be
    \frac{d\ln L_{\rm tail}}{d\ln t}\approx 1+\lp \frac{t'}{t}\rp^2.
\ee
For $t\ll t'$, this shows that $L_{\rm tail}(t)$ can become
rather steep, as should be expected because of the exponential
dependence it has on time. For $t\approx t'$, the slope of $L_{\rm
tail}(t)$ is at its shallowest with roughly $L_{\rm tail}\propto
t^2$. From this we conclude that a rising lightcurve
that is shallower than $\sim t^2$ indicates that direct heating from
$^{56}$Ni is dominating, as shown in the upper panel of Figure
\ref{fig:diagram2}. In such a case, $M_{56}$ can be approximated
from the observations and $X_{56}\approx M_{56}/\Delta M_{\rm diff}$
can be inferred as a function of time if the explosion time is well
constrained. Otherwise, a rising lightcurve steeper than $\sim t^2$
indicates that the diffusive tail is having an impact as shown in
the bottom panel of Figure \ref{fig:diagram2}, and this diffusive
tail must be accounted for in order to derive $X_{56}$.

This conclusion is only approximate since we are focusing on the the
diffusive tail from a single depth. This is likely not too bad of a simplification,
since the diffusive tails from $^{56}$Ni at even
larger depths are exponentially suppressed. In detail though, the
luminosity of the diffusive tail should depend on the sum of
contributions from all depths. We calculate and discuss the impact
of this in Appendix \ref{sec:tail}. The conclusion is that even in
instances when local radioactive heating dominates, the expected
luminosity given by equation (\ref{eq:luminosity56}) is never more
accurate than a factor of $\sim2$.

\subsection{The Importance of the Time of Explosion}

Our discussion thus far makes it clear that it is very important to know the time of explosion. It sets the time at which the thermal diffusion wave begins backing its way into the expanding ejecta, and from this at all later times it is roughly known what depths of the exploding star are being probed by the observations via equation (\ref{eq:mdiff}).

When the time of explosion is not known from direct detection of shock breakout or shock heating of the surface layers, our discussion of the early lightcurve should also provide some reason for caution. To illustrate the problem, in the top and middle panels of \mbox{Figure \ref{fig:diagram}} we compare the lightcurves for different $^{56}$Ni depositions. In the top panel, the $^{56}$Ni is deposited into rather shallow layers, therefore the timescale between the beginning of the explosion and the rise of the lightcurve is fairly short. In this case, the time of explosion could be reasonably well-approximated by extrapolating the $^{56}$Ni lightcurve back in time. In the middle panel, the $^{56}$Ni is deposited in deeper layers, and correspondingly the delay between the shock heating and rising lightcurve is longer. In this case, extrapolating the $^{56}$Ni lightcurve back in time would provide a poor estimate for the time of explosion. We are lead to the following important conclusion: {\em one cannot simply estimate the time of explosion by extrapolating the rising lightcurve back in time because its position relative to the moment of explosion depends on the depth of radioactive heating.} This means that for events like SN 2011fe, the constraints on $R_*$ cannot be as tight as previously reported \citep{Bloometal2012} when the explosion time is not known.

The earliest detection of the rising lightcurve from $^{56}$Ni does not probe the shallowest layers of the star, but merely the shallowest deposits of radioactive heating. For example if the diffusive tail only reaches $^{56}$Ni after traveling $\sim0.1M_\odot$ below the exploding star's surface, then it will take $\sim2\ {\rm days}$ to detect this depth (using eq. [\ref{eq:mdiff}]). Depending on the specific parameters for a given SN, this timescale can even be a few times larger. Similar delays between the moment of explosion and the rising lightcurve are seen in the numerical work of \citet{Dessartetal2011}.

\subsection{Clues about the Depth of $^{56}$Ni}
\label{sec:clues}

When the shock breakout and shock heating are not detected,
additional information from color, or even better
spectroscopic, observations can be used to break the degeneracy
between the depth of $^{56}$Ni and the time of explosion. We
summarize some of the properties of SNe that are most useful for
doing this in this section and the next.

In the  bottom panel of Figure \ref{fig:diagram} we schematically
show the time-dependent radius of the photosphere (orange curve)
during a SN. For a polytropic index of $n=3$, the photospheric
radius is \citep{RabinakWaxman2011},
\be
    r_{\rm ph}(t) \approx 3\times10^{14}\frac{\kappa_{0.1}^{0.11}E_{51}^{0.39}}{M_1^{0.28}}
    \lp \frac{t}{1\ {\rm day}}\rp^{0.78}\ {\rm cm},
    \label{eq:rph}
\ee
where we have suppressed the  dependence on the density
structure factor $f_\rho$ and set it to $0.01$ (see Appendix
\ref{sec:structurefactor}). Even though $r_{\rm ph}$ is moving into the star as the ejecta
expands, it is always moving out in an Eulerian frame. The observed color
temperature is
\be
    T_c \approx \lp \frac{L\tau_c}{4\pi r_{\rm c}^2\sigma_{\rm SB}}\rp^{1/4}\gtrsim \lp \frac{L}{4\pi r_{\rm ph}^2\sigma_{\rm SB}}\rp^{1/4},
    \label{eq:tc}
\ee where $r_c$ is the color radius and $\tau_c$ is the optical
depth at the color (thermalization) depth. The inequality comes from
the fact that $\tau_c \geq 1$ and $r_c \leq r_{\rm
ph}$. For typical parameters $\tau_c$ is not much larger than unity
(see \mbox{Appendix \ref{sec:Thermalizationdepth}}) and so is
$r_{\rm ph}/r_c$, so this inequality is also a fair approximation of
$T_c$. Although $r_{\rm ph}$ roughly always scales as a power-law
with time (eq. [\ref{eq:rph}]), $L(t)$ can vary greatly depending on
the $^{56}$Ni deposition. Thus, the color evolution
during the rising phase provides the first clue on the $^{56}$Ni
depth. For deep $^{56}$Ni, the initial rising phase of $L$ is
exponential, so from equation (\ref{eq:tc}) we see that $T_c$ should
be rising as well. On the other hand, if $^{56}$Ni is shallow then
$L$ rises more gradually and $T_c$ is roughly constant or even
slowly decreasing during the lightcurve rise.

In the bottom panel of Figure \ref{fig:diagram} we also show the photospheric velocity (green curve). Taking $v_{\rm ph}=r_{\rm ph}/t$, we find
\be
    v_{\rm ph}(t)\approx 35,000\frac{\kappa_{0.1}^{0.11}E_{51}^{0.39}}{M_1^{0.28}}
    \lp \frac{t}{1\ {\rm day}}\rp^{-0.22}\ {\rm km\ s^{-1}}.
    \label{eq:vph}
\ee
Unlike the photospheric radius, the photospheric velocity is decreasing with time. This is because the photosphere is moving into the star in a Lagrangian sense, and reaching material that is moving at successively lower velocities. Furthermore, the time derivative of the photospheric velocity is $dv_{\rm ph}/dt\propto t^{-1.22}$. Therefore we expect the change in the photospheric velocity to become more gradual with time.

Using the above arguments, we conclude that there are four main ways in which the depth of $^{56}$Ni heating may be qualitatively inferred. All other things being equal, a larger $^{56}$Ni depth implies the following.
\begin{enumerate}
\item The photospheric radius is larger when the $^{56}$Ni heating gets out, and therefore the temperature will be lower.
\item The luminosity at early times is increasing faster than the radius expands, causing temperature to evolve bluer.
\item The photospheric velocities are lower.
\item The photospheric velocities evolve more slowly with time.
\end{enumerate}

\subsection{The Minimum Explosion Time}

In additional to the qualitative correlations listed above, a
single measurement of the photosphere velocity and
color temperature during the rise can provide a strict, model
independent, upper limit to the time of
explosion before that measurement. This is seen by first rewriting
equation (\ref{eq:tc}) as
\be
    L \approx 4\pi r_c^2 \sigma_{\rm SB} T_c^4/\tau_c,
\ee
and then setting $r_c \approx v_ct_{\rm exp}$ where $t_{\rm
exp}$ is the time since the explosion began, to estimate that
\be
    t_{\rm exp} \approx \lp \frac{L\tau_c}{4\pi v_c^2\sigma_{\rm SB}T_c^4} \rp^{1/2}.
    \label{eq:t}
\ee
In general $T_c$ can be measured directly from the observations,
but any velocity that can be measured will be from an absorption
feature shallower than the photosphere and generally $v_{\rm
ph}\gtrsim v_c$. Furthermore, $\tau_c$ will be greater than unity
but also difficult to infer just from the observation. It is
therefore useful to have a quantity that can simply be estimated
directly in terms of observable quantities,
\be
    t_{\rm min} \equiv \lp \frac{L}{4\pi v_{\rm ph}^2\sigma_{\rm SB}T_c^4} \rp^{1/2} = \frac{t_{\rm exp}}{\tau_c^{1/2}}\lp\frac{v_c}{v_{\rm ph}} \rp
    \lesssim t_{\rm exp}.
\ee
Substituting typical values we find
\be
    t_{\rm min} = 4.3 \lp \frac{L}{10^{42}\rm\ {erg\ s^{-1} }}\rp^{1/2}
    \lp \frac{T_c}{10^4\rm{~K}} \rp^{-2}
    \nonumber
    \\
    \times\lp \frac{v_{\rm ph}}{10^4 {\rm ~km\ s^{-1}}}\rp^{-1}{\rm days}.
    \label{eq:tmin}
\ee
Therefore $t_{\rm min}$ is an {\em observable quantity that provides
a model independent lower limit to the time of explosion}\footnote{In \mbox{Appendix
\ref{sec:Thermalizationdepth}} we derive the value of $\tau_c$ and show
that it is not much larger than unity. For example, even for high
temperature of $20,000$ K and $v_{\rm ph}=20,000$ km/s at 10 days after the explosion
$1<\tau_c \lesssim 6$, implying $t_{\rm min}<t_{\rm exp}\lesssim 2.5 t_{\rm min}$.
For lower temperature and velocity the range is even smaller. Therefore $t_{\rm min}$
is not only a lower limit but also a rough approximation of $t_{\rm exp}$.}.
Another reason that
$t_{\rm min}$ is especially useful is that it only requires that the
velocity and temperature of the SN be obtained at a {\em single
time}. Thus when resources are limited, using equation
(\ref{eq:tmin}) is a helpful technique for learning a lot about the
SN with minimal additional investment.

This concludes our discussion of the qualitative features of rising
SN I lightcurves. In general specific events may
have details that we have not addressed, like non-spherical
explosions or complicated velocity profiles. But by clearly spelling
out the chain of reasoning behind our conclusions we provide rules
of thumb that can be used to build intuition, even in cases that are
somewhat more complicated. To illustrate how our arguments can be
used in practice, in the next section we apply them to a
recently-discovered, well-studied SNe Ic.

\section{PTF 10vgv}
\label{sec:10vgv}

PTF 10vgv is a SN Ic that was discovered on 2010 September 14.1446
(UTC time) with the Palomar Oschin Schmidt 48 inch telescope (P48),
by the PTF survey \citep{Corsietal2012}. In a previous image taken
on 2010 September 12.4830, it was not seem down to a limiting
magnitude of $R>20.2$. Following detection, the $R$-band
luminosity rises quickly to a peak $\approx10$ days later.
A single spectrum was taken $\approx2$ days after
detection. In light of what we have been discussing, it would have
been ideal if the emission from shock heating could have been
identified. It would have provided a measurement of
the progenitor radius and the time of explosion, which could have
been used to directly probe the $^{56}$Ni deposition during the
rising lightcurve. Although it is unfortunate that this was not available, this event is ideal as a test case for
exploring what can be learned when the time of explosion is
not known.

\subsection{Radius Constraints}
\label{sec:radius}

 \begin{figure}
\epsscale{1.2}
\plotone{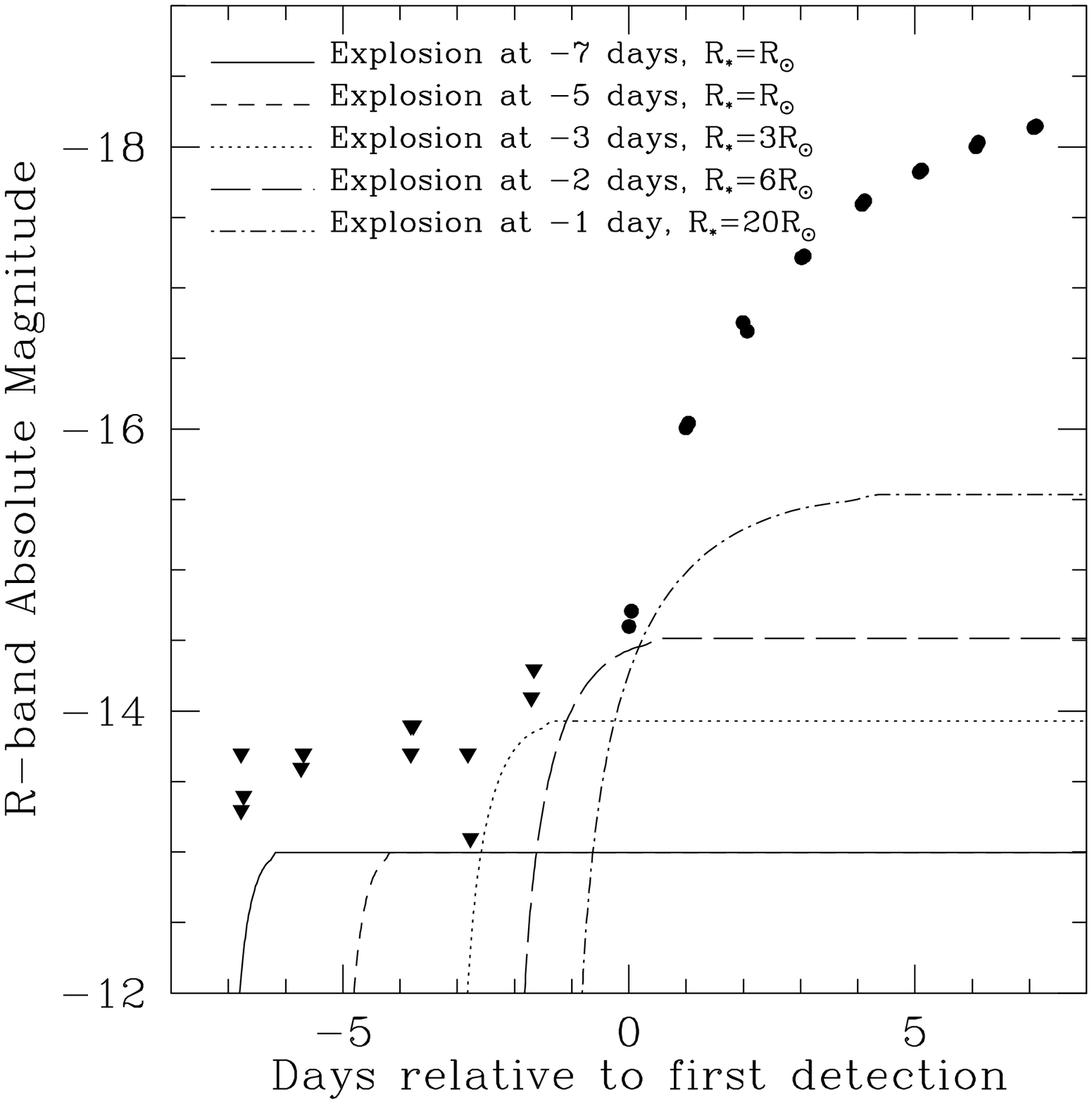}
\caption{The circles and triangles plot the data for PTF 10vgv from \citet{Corsietal2012} for the detections and upper limits, respectively. The curves show theoretical calculations of the shock heating lightcurves using equations (\ref{eq:L_coolingPhase}) and (\ref{eq:T_coolingPhase}), which are set to plateau at a temperature of $\approx0.6\ {\rm eV}$, for a range of explosion times and radii as labeled. In all cases we fix \mbox{$E=10^{51}\ {\rm erg}$} and \mbox{$M=2M_\odot$} for the sake of comparison. The main conclusion is that without direct identification of the time of explosion, the radius upper limit constraint can vary by an order of magnitude or more.}
\label{fig:shockheating}
\epsscale{1.0}
\end{figure}

Even though shock-heated cooling was not observed,
PTF 10vgv did have an early detection of the rising
lightcurve and upper limits in the time before this. Using this
information, interesting constraints on the progenitor's radius are
still possible \citep{Corsietal2012}. In Figure
\ref{fig:shockheating} we plot the observed $R$-band absolute
magnitude of PTF 10vgv (circles), along with upper limits
(triangle).  These all assume a distance modulus of 34.05 and
galactic extinction of $A_R=0.445\ {\rm mag}$ \citep{Schlegeletal1998}. We then
calculate the shock-heated cooling according to equation
(\ref{eq:L_coolingPhase}), which transitions to the plateau phase
given by equation (\ref{eq:L_plateau}). We infer the corresponding
$R$-band absolute magnitude using \mbox{equation
(\ref{eq:T_coolingPhase})}, or a plateau temperature of $\approx0.6\
{\rm eV}$, along with P48-calibrated bolometric corrections provided
by Eran O. Ofek (private communication). The resulting lightcurves
are plotted alongside the data from PTF 10vgv in Figure
\ref{fig:shockheating} for a range of explosion times and progenitor
radii. The conclusion is that without knowing the time of explosion,
the radius can only be constrained to be $R_* \lesssim
1-20R_\odot$. In general, the earlier the explosion time is, the
more stringent the constraints on the radius.

\subsection{Bolometric Lightcurve}
\label{sec:bolometric}

\begin{figure}
\epsscale{1.2}
\plotone{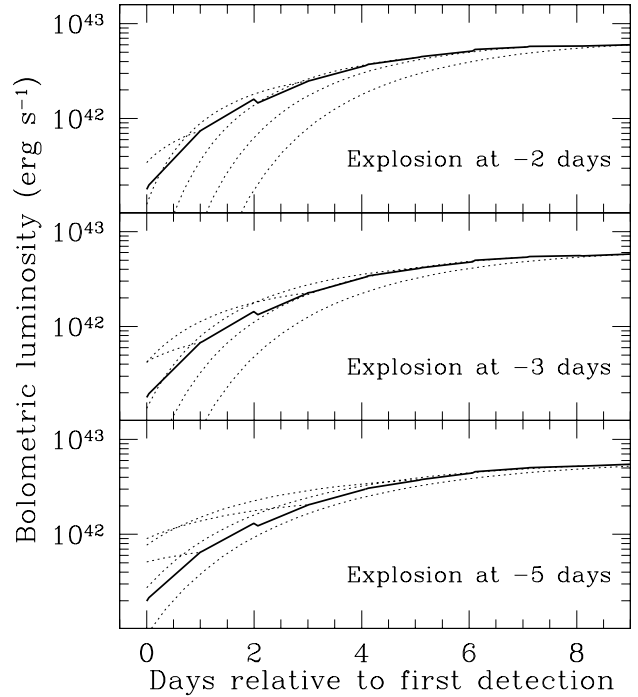}
\caption{The bolometric luminosity of PTF 10vgv (thick, solid curves). From the top to bottom panel, we consider explosion times of 2, 3 and 5 days before the first detection. In all cases we fix $E=10^{51}\ {\rm erg}$ and \mbox{$M=2M_\odot$}. For each lightcurve, at various times we consider what the diffusive tail coming from that depth would look like at earlier times (dotted lines) using equation (\ref{eq:tail}). From this comparison we conclude that if the SN occurred recently (top panel), then only the earliest heating may be explained by a diffusive tail and the majority of the rising lightcurve directly probes the $^{56}$Ni distribution. On the other hand, if the explosion occurred further in the past (the middle and bottom panels), then the majority of the rising lightcurve can be explained as a diffusive tail due to deep deposits of $^{56}$Ni. This example demonstrates the degeneracy between the time of explosion and the depth of $^{56}$Ni if only a photometric lightcurve is available.}
\label{fig:lum_diff}
\epsscale{1.0}
\end{figure}

Next we consider what can be learned about the mass and distribution
of $^{56}$Ni from the observed lightcurve. Before this can be done,
there are two considerations that must be made. First, we need to
convert the observed $R$-band magnitudes to a bolometric luminosity.
We solve for this using
\be
    L \approx 4\pi r_c^2 \sigma_{\rm SB}T_c^4/\tau_c\approx10^{(M_{R,\odot}-M_R-BC)/2.5}L_\odot.
    \label{eq:l}
\ee
Since $BC$ depends on $T_c$, we can self-consistently find
$T_c$ and thus $L$ at any given time. For our estimates here we
simply take $\tau_c\sim1$ and $r_c\sim r_{\rm ph}$ as given by
equation (\ref{eq:rph}) to produce the thick, solid curves in Figure
\ref{fig:lum_diff}. Although not physically accurate, these
estimates are adequate since the bolometric luminosity is found to
be fairly robust. Once again, we must take into account that the
time of explosion is not known, so we consider three example
explosion times.

The second consideration we must make before attempting to associate
the bolometric lightcurve with the $^{56}$Ni distribution is the
impact of a diffusive tail. Assuming that the bolometric lightcurves
in Figure \ref{fig:lum_diff} are representative of the $^{56}$Ni
distribution, we can then ask, for a given luminosity at a given
time what is the implied diffusive tail at
earlier times? The dotted lines in Figure \ref{fig:lum_diff} show
the diffusive tails originating from various times using equation
(\ref{eq:tail}). If the dotted line sits below the solid curve, this
means that the diffusive tail is insufficient to explain the
bolometric lightcurve at this depth, and thus the bolometric
lightcurve represents direct heating from $^{56}$Ni (as shown in the
top diagram in Figure \ref{fig:diagram2}). On the other hand, if the
dotted line sits above the solid curve, then the diffusive tail from
that point overestimates the earlier bolometric lightcurve and
luminosity at that time cannot be from direct heating of
$^{56}$Ni at the diffusion depth (as shown in the bottom diagram in
Figure \ref{fig:diagram2}).

The conclusion
from this comparison is that it is difficult to determine which of
these three explosion times are more accurate from just this
information. If the SN occurred recently (top panel of Figure
\ref{fig:lum_diff}), then only the earliest heating is explained
by a diffusive tail and we would infer that the majority of the
rising lightcurve directly probes the $^{56}$Ni distribution. On the
other hand, if the explosion occurred further in the past (the
middle and bottom panels of Figure \ref{fig:lum_diff}), then the
majority of the rising lightcurve is explained as a diffusive
tail due to deep deposits of $^{56}$Ni. This is consistent with our
discussion in \S \ref{sec:clues} that there is a degeneracy between
these two limits unless more information is available.

\begin{figure}
\epsscale{1.2}
\plotone{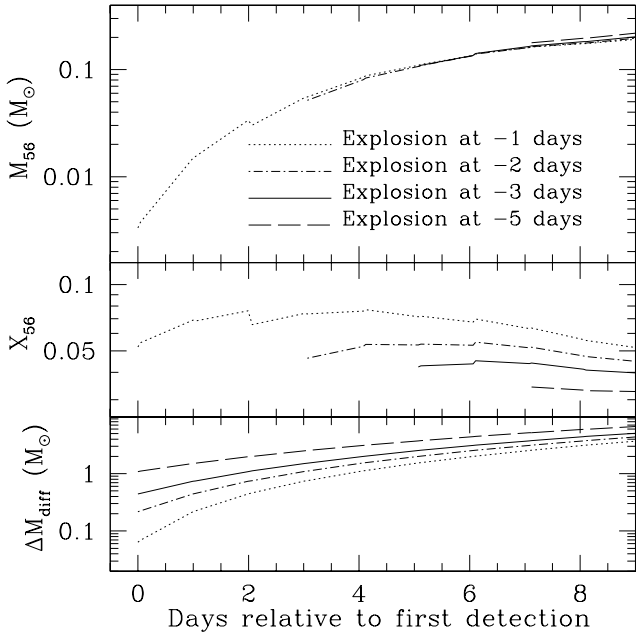}
\caption{Inferred mass, mass fraction, and depth of $^{56}$Ni in PTF 10vgv for different explosion times. The $^{56}$Ni mass $M_{56}$ is inferred from \mbox{equation (\ref{eq:luminosity56}),} and we only plot $M_{56}$ and $X_{56}$ for times at which the bolometric luminosity cannot be explained as a diffusive tail according to Figure \ref{fig:lum_diff}. The mass fraction is estimated as $X_{56}\approx M_{56}/\Delta M_{\rm diff}$, where the diffusion depth is calculated according to equation (\ref{eq:mdiff}).}
\label{fig:ni56}
\epsscale{1.0}
\end{figure}

Nevertheless, we can still try to constrain the mass and distribution of $^{56}$Ni as a function of the explosion time, the results of which are shown in Figure \ref{fig:ni56}. In the top panel we plot the mass of $^{56}$Ni inferred from equation (\ref{eq:luminosity56}). This is only plotted for depths at which the bolometric luminosity cannot be explained by a diffusive tail using Figure \ref{fig:lum_diff}. In the middle panel we plot the mass fraction $X_{56}$, and in bottom panel the thermal diffusion depth using equation (\ref{eq:mdiff}).

\subsection{Temperature and Velocity Constraints}

We next consider what additional information can be learned about
PTF 10vgv from temperature and velocity measurements. Unfortunately,
the available data during the rise is rather limited.
\citet{Corsietal2012} mention a $16,000\ {\rm km\ s^{-1}}$ Si II
absorption feature \citep[which is typically associated with the
photosphere,][]{Tanakaetal2008} observed on 2010 September 16, which
corresponds to $\approx2\ {\rm days}$ after discovery. From the
spectrum taken on that same day, by eye one can see that it roughly
peaks at \mbox{$\sim4300$ \AA,}  which,
assuming a blackbody spectrum, corresponds to a temperature of
$\sim6700\ {\rm K}$. But we must be cautious. In the study of SN
1994I by \citet{Saueretal2006}, a spectrum with a similar peak is
observed at $8\ {\rm days}$ post explosion, and their detailed
models infer a temperature of $\sim10,000\ {\rm K}$. As we
describe next, if the correct temperature were known, then tight
constraints could be placed on the time of explosion and thus the
other properties of PTF 10vgv. But given that the temperature
cannot be extracted from the data without a detailed spectral
modeling, which is beyond the scope of this paper, we separately
consider both the cases of $6700\ {\rm K}$ and $10,000\ {\rm K}$ and
discuss the implications.

Using equation (\ref{eq:tmin}) we estimate the minimum time of
the explosion. For $T_c=6700\ {\rm K}$, $v_{\rm ph}=16,000\ {\rm km\
s^{-1}}$, and $L\approx1.5\times10^{42}\ {\rm erg\ s^{-1}}$ (taken
from Figure \ref{fig:lum_diff}), we find $t_{\rm min}\sim7\ {\rm
days}$. Therefore, if the cooler temperature is the correct one
then the explosion must have occurred $\sim5\ {\rm days}$ or more
before the first detection. On the other hand, for $T_c=10,000\ {\rm
K}$ we estimate $t_{\rm min}\sim3\ {\rm days}$ and the explosion
must have occurred $\sim1\ {\rm day}$ or more before the first
detection.

\begin{figure}
\epsscale{1.2}
\plotone{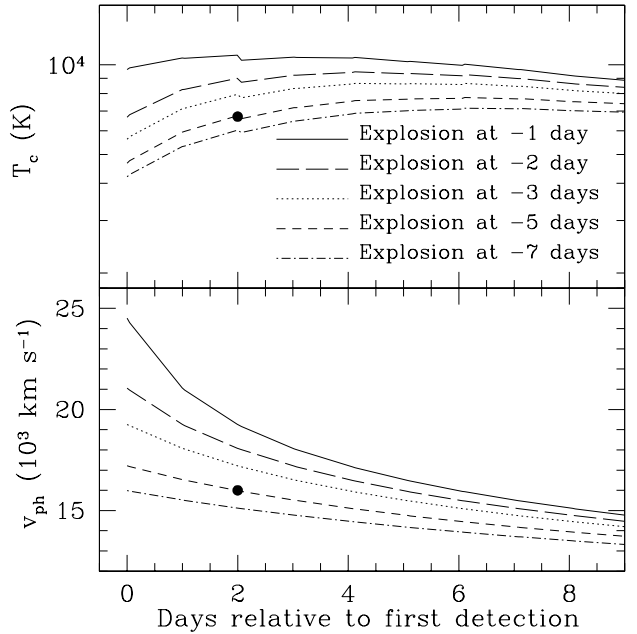}
\caption{Temperature and velocity evolution for PTF 10vgv when fitting a temperature of $6700\ {\rm K}$ and velocity of $16,000\ {\rm km\ s^{-1}}$ at 2 days past first detection (shown as filled circles). In this particular case, the best fits occur for $E_{51}=0.41M_1^{0.72}$. The curves are theoretical calculations assuming different explosion times as labeled, which demonstrates than an explosion time of $\sim5\ {\rm days}$ before first detection is favored.}
\label{fig:t_and_v}
\epsscale{1.0}
\end{figure}

\begin{figure}
\epsscale{1.2}
\plotone{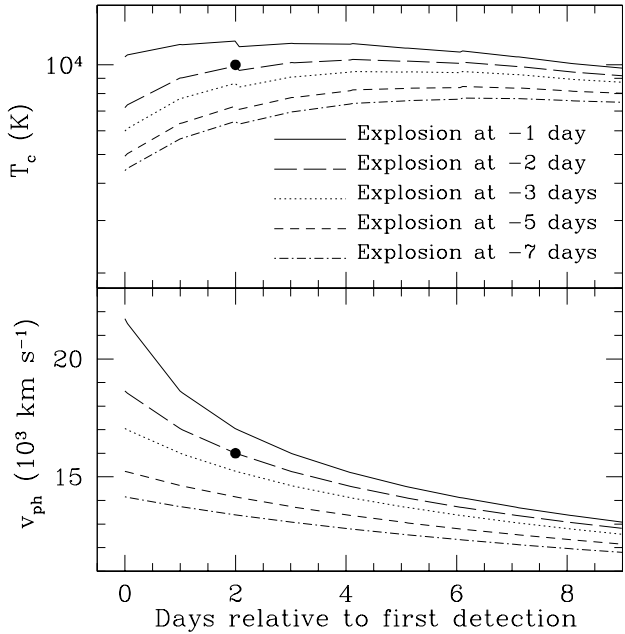}
\caption{The same as Figure \ref{fig:t_and_v}, but this time fitting a temperature of $10,000\ {\rm K}$. Now the best fits occur for $E_{51}=0.30M_1^{0.72}$. In this hotter case, the explosion time is constrained to be $\sim2$ days before first detection.}
\label{fig:t_and_v2}
\epsscale{1.0}
\end{figure}

To test these conclusions, we perform a detailed fit of the
temperature, velocity, and bolometric luminosity constraints
simultaneously. This is done by varying $E$, $M$, and the time of
explosion over a wide range of values and identifying which best fit
the constraints. The result for either $6700\ {\rm K}$ or $10,000\
{\rm K}$ is that a fit can only be obtained when $E \propto
M^{0.72}$ because this fixes the normalization of the velocity (see eq. [\ref{eq:vph}]). But
for a given temperature only a very narrow range of explosion times
are consistent with the data. In Figures
\ref{fig:t_and_v} and \ref{fig:t_and_v2} we plot the results of our
fitting. These show that for a temperature of $6700\ {\rm K}$ or
$10,000\ {\rm K}$, the explosion much have occurred $\sim5\ {\rm
days}$ or $\sim2\ {\rm days}$ prior to first detection,
respectively. These match the values of $t_{\rm min}$ estimated
before.

Besides showing how the explosion time can be constrained, Figures
\ref{fig:t_and_v} and \ref{fig:t_and_v2} highlight many of the
general trends that we discussed in \S \ref{sec:summary} when
considering how different explosion times impact various observables
of the SN. For a more recent explosion time, the gradient in
$v_{\rm ph}$ is greater and the temperature is higher and
decreasing. In contrast, for an explosion time more distant in the
past, $T_c$ is lower and increasing for a longer time. If
merely a couple of data points were available, these trends could be
identified in the observations, and even tighter constraints could
be placed on PTF 10vgv. Nevertheless, it is powerful that even a
{\em single} velocity and temperature measurement provides such
stringent constraints. SNe Ia tend to have more early data available
than SNe Ib/c, and we consider some of these events
in forthcoming work.

\subsection{Progenitors}

Our conclusions from the previous sections show just how greatly our inferences about the progenitor can vary depending on the explosion time. These are summarized as follows.
\begin{enumerate}
\item If the temperature at 2 days past first detection is \mbox{$\sim6700\ {\rm K}$,} then the explosion occurred $\sim 5\ {\rm days}$ or more before first detection. The radius is constrained to be $\lesssim1R_\odot$ and the $^{56}$Ni must be located much deeper in the ejecta. Also the ejecta is likely of higher mass since it takes longer to reach the SN peak.
\item If the temperature at 2 days past first detection is \mbox{$\sim10,000\ {\rm K}$,} then the explosion occurred merely \mbox{$\sim 2\ {\rm days}$} before first detection. The radius is constrained to be
$\lesssim6R_\odot$ and the $^{56}$Ni must be located fairly
shallowly. Also the ejecta is likely of lower mass.
\end{enumerate}
So we are left with two seemingly opposite conclusions about almost
all the aspects of the ejecta just from opposite assumptions about
the time of explosion. We hope that a proper modeling of the
available spectrum, which we do not carry out here, can provide the
correct temperature and decide between these two options.
We next discuss the implications for the progenitor star that
produced PTF 10vgv in each case.

Since the progenitors of such SN Ic are
hydrogen-stripped stars, the models of \citet{Yoonetal2010} are a
helpful starting place. The most striking
feature of these progenitors is the inverse relationship between
mass and radius that is shown in their Figure 12. If PTF 10vgv
had a radius $\lesssim1R_\odot$, it is strongly inconsistent with any
of the binary models below $M_f\lesssim 7M_\odot$, where $M_f$ is
the final mass at the time of explosion, which all have radii larger than
$\sim2R_\odot$. Their helium star models (i.e., Wolf-Rayet
progenitors)  have smaller radii and may be consistent with
$M_f\gtrsim6-8M_\odot$, depending of the mass loss prescription
employed\footnote{In previous calculations of helium stars, such as
by \citet{Woosleyetal1995}, smaller radii are generally seen. This
difference is thought to be due to both different mass loss rates
and updated opacities; \citet{Yoonetal2010} use opacities from
\citet{IglesiasRogers1996}.}. Other progenitors that are consistent
with such small radii are the gamma-ray burst models of
\citet{Yoonetal2006} with $M_f\gtrsim 12M_\odot$, but these probably
are not applicable to PTF 10vgv.

Combining the generally large mass inferred for a small-radius progenitor
with an ejecta mass of $\sim3-5M_\odot$ (estimated
from $\Delta M_{\rm diff}$ at peak), the remnant mass
must be $\gtrsim2-3M_\odot$. This is near the range of the
maximum mass of normal NSs \citep{LattimerPrakash2001}. Would this
imply that this event led to the formation of a black hole? Given the
number of estimates that have been made throughout our analysis, it
is difficult to make this conclusion with such certainty. But this
does demonstrate how our analysis can make interesting connections
between observations on one hand and the
detailed simulations of progenitors on the other.

For the case where
$R_*\lesssim6R_\odot$, \citet{Yoonetal2010} show that such radii are naturally expected for
progenitors in a mass range of $M_f\sim4-5M_\odot$, and it is easier to
reconcile the ejecta mass estimate with formation of a neutron star
remnant. A less massive progenitor with a large
radius has been predicted to have mixing of $^{56}$Ni into the outer
layers by Rayleigh-Taylor instability during the supernova explosion
\citep{Hachisuetal1991,Joggerstetal2009}, similar to what we infer for a recent
explosion (Fig. \ref{fig:ni56}). So even in this case there
are interesting correlations and implications for the explosion mechanism that
can be investigated.

Finally, it is worth discussing the inferences that can be made from
the fact that PTF 10vgv was classified as a SN Ic.
\citet{Dessartetal2012} studied the impact of $^{56}$Ni mixing on
the ejecta of SNe Ib/c, and found that it strongly impacts the
observed spectral features. Chief among these are the \mbox{He I}
lines, which require non-thermal electrons that are excited by
$\gamma$-ray lines from $^{56}$Ni and $^{56}$Co when
the helium mass fraction is $\lesssim0.5$ \citep{Dessartetal2011}.
This means that the same ejecta with a significant amount of
helium can produce either a SN Ib if the $^{56}$Ni maxing is strong
or a SN Ic if the $^{56}$Ni mixing is weak. If PTF 10vgv
were constrained to have a recent explosion with strong
$^{56}$Ni mixing, we would be forced to conclude that its outer
layers are very helium-poor. In the future it would be informative
to repeat our analysis on a SN Ib to see if shallow $^{56}$Ni can be
inferred to make the He I visible.

\section{Conclusions and Discussion}
\label{sec:conclusion}

We have discussed the photometric rising lightcurves of SNe I to
investigate what can and cannot be learned from these observations.
We provided a detailed summary of the various stages that
determine the early lightcurve, and how their relative contributions
are impacted by $^{56}$Ni mixing and uncertainties in the explosion
time. We then looked at a particular SN Ic event PTF 10vgv in some
detail. Even with no direct constraint on the explosion time, and
from just an $R$-band photometric lightcurve and a single estimate
of the velocity and temperature of the ejecta, we argued that the
explosion time could be constrained, which would have had a number of
implications for the progenitor that we discussed.

Our investigation demonstrates the kinds of connections that can be
made between observations, explosion calculations, and progenitor
models. To facilitate such connections in
the future, we summarize our conclusions below. We also
highlight observational information that we deem especially
important, and which should be considered of high priority, when
investigating a rising lightcurve on limited resources.
\begin{enumerate}
\item A SN may exhibit a dark phase between the moment of explosion and the rise of the $^{56}$Ni lightcurve. This means that extrapolating the $^{56}$Ni lightcurve back in time is not a reliable method for estimating the time of explosion, and that without a known explosion time, constraints on $R_*$ \citep[such as for SN 2011fe;][]{Bloometal2012} are less stringent.
\item Even though shock breakout may only emit at short wavelengths
and may be too short-lived for detection in most circumstances, the
UV/optical detection of shock-heated cooling phase can be just as
useful for putting constraints on the progenitor radius.
\item If caught when rising (eq. [\ref{eq:F_opt_coolingPhase}]), shock-heated cooling can also identify the time of
explosion. Conversely, if the UV/optical rise is steeper than $\sim t^{1.5}$, then this argues that the shock-heated cooling is not being observed, and the explosion time is not well constrained. Due to a possible dark phase, data right before the $^{56}$Ni lightcurve may not be sufficient to catch this rise that instead may occur $\sim1-6\ {\rm days}$ earlier.
\item If the time of explosion is well constrained, then the $^{56}$Ni
mixing can be extracted from the bolometric luminosity in the
following way. First use equation (\ref{eq:luminosity56}) to find
$M_{56}(t)$ assuming there is no contribution from the diffusive
tail. Then at any time $t'$ find the diffusive tail contribution from
$M_{56}(t)$ at $t<t'$ (eq. [\ref{eq:tail}]). If this
contribution falls below the actual luminosity then $M_{56}(t)$ is a
reliable estimate (to within a factor of $\approx2$) of the $^{56}$Ni mass that
mixed into $\Delta M_{\rm diff}(t)$. Otherwise it is not. An upper limit
on $M_{56}(t)$ can be derived then by finding the largest $^{56}$Ni
mass that produces a diffusive tail that falls below the observed
luminosity. A useful rule of thumb is that if the luminosity
rises faster than $\sim t^2$, then it is dominated by the diffusive
tail.
\item If the time of explosion is unknown, having even a {\em single} temperature and velocity measurement during the rise can go a long way toward supplementing the photometric data and provide strong constraints on the time of explosion using equation (\ref{eq:tmin}).
\item Having $\gtrsim2$ temperature and velocity measurements during the rise will allow the time evolution of each to be constrained in more detail. Knowing the time evolution of the temperature is especially useful because it provides a check of any conclusions about $^{56}$Ni mixing.
\item A future goal of observations should be to obtain similar coverage of SNe Ib/c with strong $^{56}$Ni mixing. Such events would likely be classified as SNe Ib unless completely devoid of helium \citep{Dessartetal2012} and should never show a shock-heated cooling phase. This would be an important test of whether all the telltale signs of $^{56}$Ni mixing are present as we have outlined.
\end{enumerate}

Although the scalings we have used are
fairly robust, the numerical prefactors come from
semi-analytic work that is only approximately correct. A useful
future project would therefore be to calibrate our prefactors against
more detailed numerical calculations, in particular the photospheric
velocity, color temperature, and diffusion depth. As highlighted by \citet{Dessartetal2012}, the opacities
for these ejecta can become complicated, so the various fits may
only be applicable over different subsets of parameter space in
composition and temperature. Nevertheless, the resulting collection of
relations would be useful tools for quickly estimating properties of
the explosions and the progenitors for future observations.
\\

\acknowledgments
We thank Eran Ofek for assistance in implementing
the bolometric corrections for the P48, Avishay Gal-Yam for providing detailed
comments, and Alessandra Corsi for
helping with information about PTF 10vgv. We also thank Luc Dessart,
Peter Goldreich, Christian Ott, and Re'em Sari for thoughtful
discussions.
A.L.P. was supported through NSF grants AST-1212170,
PHY-1151197, and PHY-1068881, NASA ATP grant NNX11AC37G,
NSF grant AST-0855535, and by the Sherman Fairchild Foundation.
E.N. was partially
supported by an ERC starting grant (GRB-SN 279369).

\begin{appendix}

\section{Shock-heated Cooling in the Rayleigh-Jeans Tail}
\label{sec:rayleighjeanstail}

At early times, when the lightcurve is dominated by shock-heated
cooling of ejecta and before $^{56}$Ni heating has become important,
the lightcurve can rise in the optical/UV even if the bolometric
luminosity is declining. This is possible as long as the optical/UV
emission is in the Rayleigh-Jeans tail. In this limit \be
    L_{\rm opt/UV} \propto \frac{r_c^2T_c}{\tau_c},
\ee where $r_c$ is the color radius and $\tau_c$ is the optical
depth at the color depth. Using the scalings from equations (20) and
(21) of \citet{NakarSari2010}, we find that $r_c\propto t^{0.83}$
and $\tau_c\propto t^{-0.44}$ for a polytropic index of $n=3$.
Combining this with $T_c$ from equation (\ref{eq:T_coolingPhase})
above, this gives $L_{\rm opt/UV}\propto t^{1.5}$ as summarized in
\S \ref{sec:shockheating}. Therefore when a lightcurve rises faster
than $t^{1.5}$ it is strong evidence that something other than shock
heating is provided the observed energy (although this conclusion is
not foolproof, because in principle velocity gradients may differ
from $v_s\propto \rho^{-0.19}$ as we assume).

\section{The Total Contribution from the Diffusive Tail}
\label{sec:tail}

As mentioned in \S \ref{sec:shallower} and \S \ref{sec:deeper},the
total luminosity from the diffusive tail in detail is not just from
a {\em single} location. Instead it is integral over the diffusive
tails from {\em all} depths. Such an integral can be approximated as
\be\label{eq:tailtotal1}
    L_{\rm tail,total}(t)=
    \int^{t_{\rm peak}}_{t} X_{56}(t')\frac{\partial \Delta M_{\rm diff}}{\partial t'}
    \epsilon(t) \frac{{\rm erfc}( t'/\sqrt{2}t )}{ {\rm erfc}(1/\sqrt{2} )} dt',
\ee where we take the limit of the integration to be the time of
peak $t_{\rm peak}$ since this roughly corresponds to when the
diffusion wave has made its way completely through the ejecta.
Making use of the time-dependence of the diffusion depth from
equation (\ref{eq:mdiff}), the integral can be rewritten as
\be\label{eq:tailtotal2}
        L_{\rm tail,total}(t)\approx 1.76 L_{56}(t)
    \int^{t_{\rm peak}}_{t} \frac{X_{56}(t')}{X_{56}(t)} \left(\frac{t'}{t}\right)^{1.76}
        \frac{{\rm erfc}( t'/\sqrt{2}t )}{ {\rm erfc}(1/\sqrt{2} )} \frac{dt'}{t'},
\ee

\begin{figure}
\centering
\begin{tabular}{cc}
\psfig{file=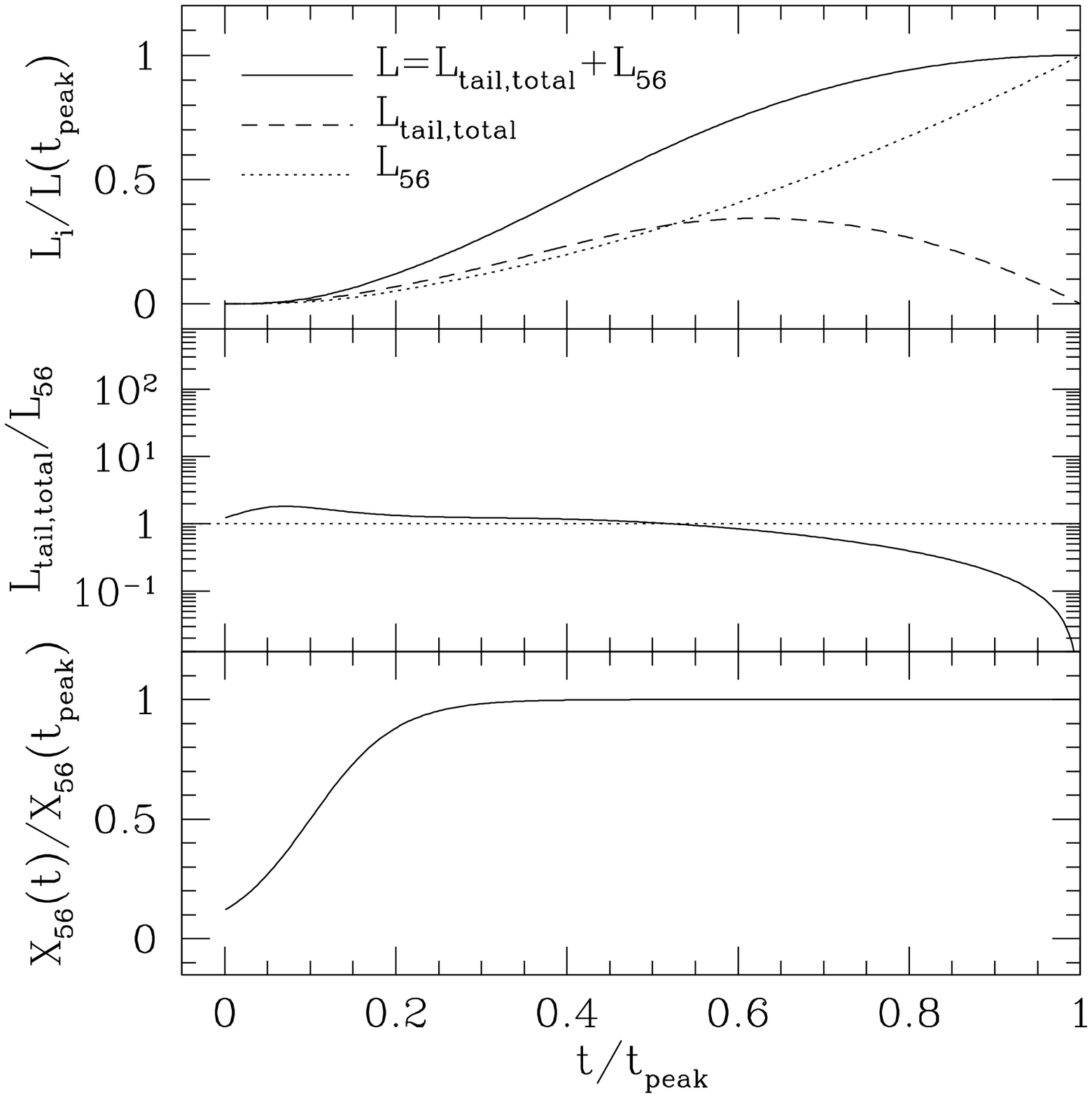, height=3.2in} &
\psfig{file=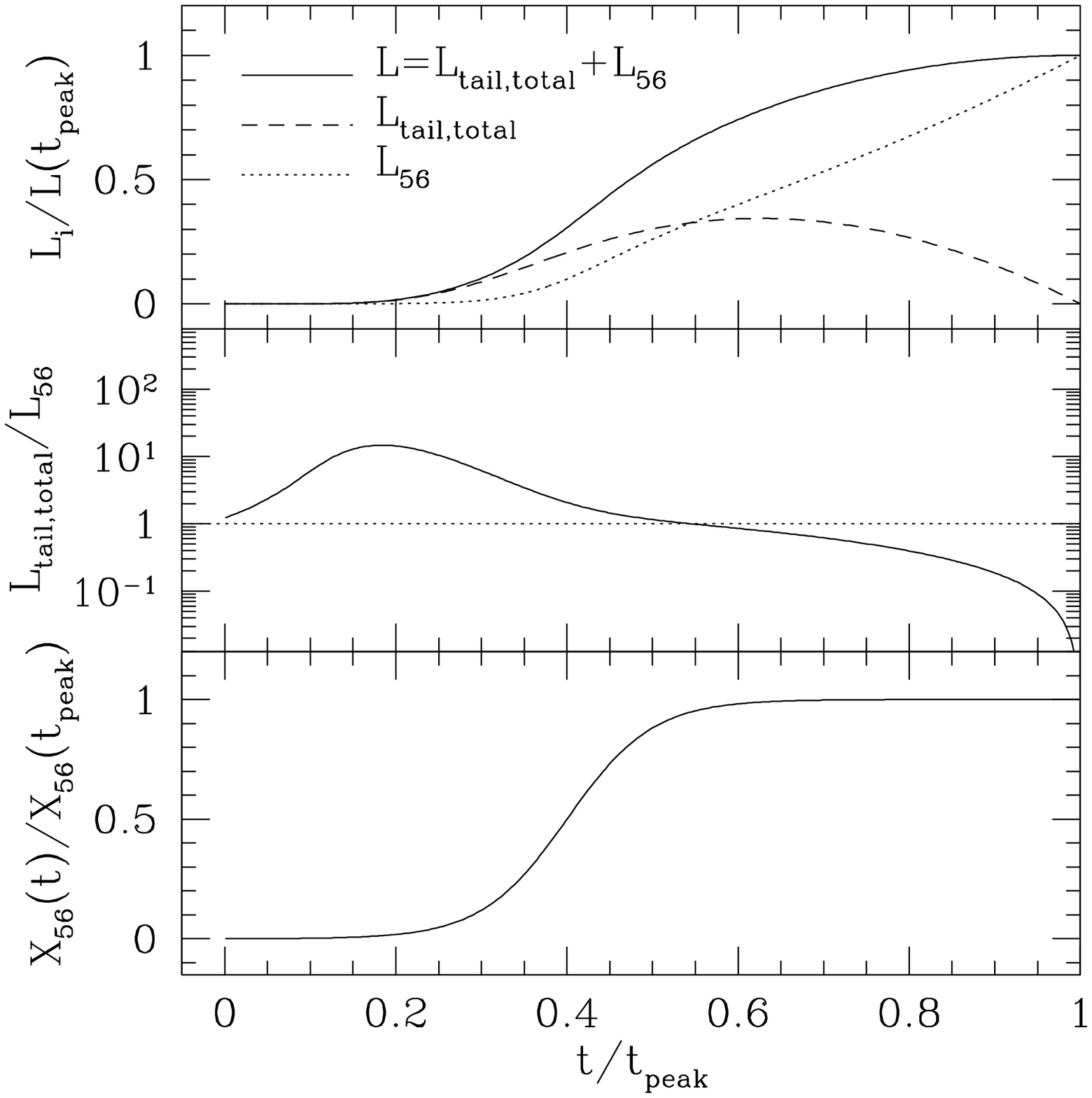, height=3.2in}\\
\psfig{file=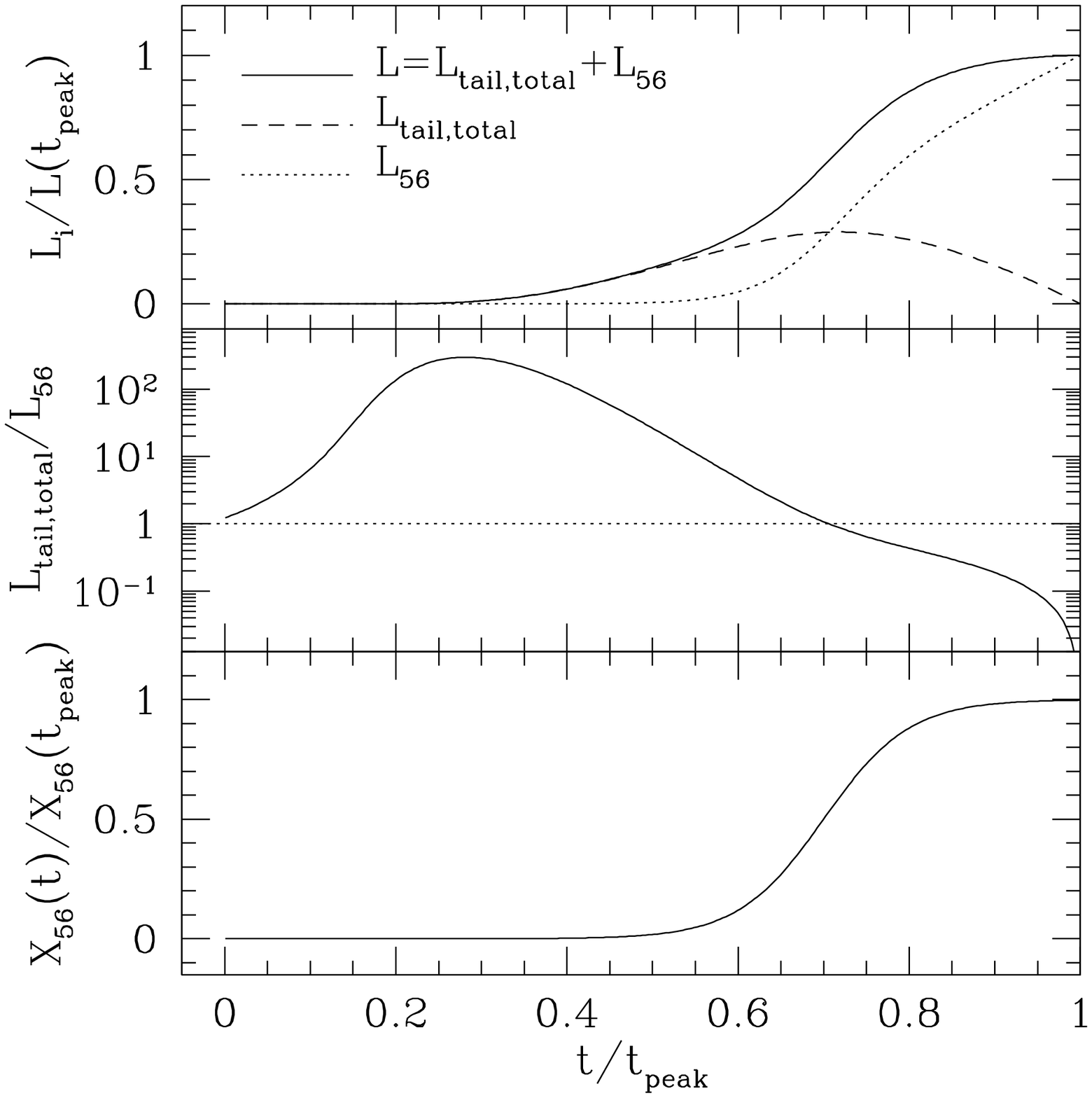, height=3.2in} &
\psfig{file=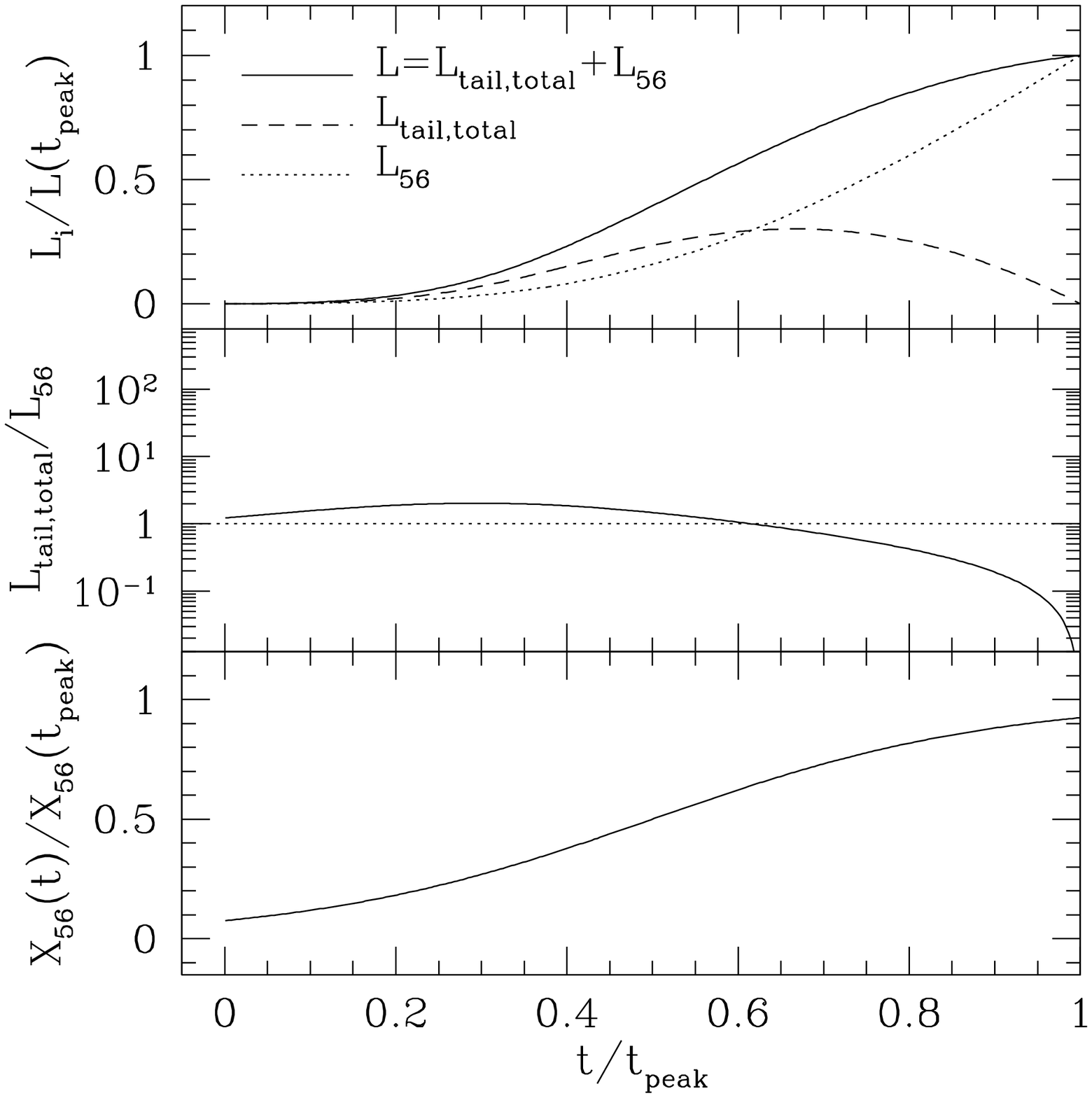, height=3.2in} \\
 \end{tabular}
\caption{Comparisons on the lightcurves (shown in each top panel)
for different distributions of $^{56}$Ni (shown in the bottom
panel), with the inclusion of the integrated diffusive tail
according to equation (\ref{eq:tailtotal2}). In the middle panels we
plot the ratio of the total tail component to the luminosity
directly from local $^{56}$Ni.} \label{fig:dt_integral}
\epsscale{1.0}
\end{figure}

In Figure \ref{fig:dt_integral} we plot the results of this integral
for various distributions of $X_{56}$. In all cases we use a
constant luminosity from $^{56}$Ni heating (no exponential decay)
for simplicity. In the first three plots we keep the shape of the
increase of $X_{56}$ the same, but vary its depth, increasing in
depth from the top left to the top right to the bottom left. The top
panel of each plot shows the various contributions to the lightcurve
as labeled. The middle panels show the ratio of the tail
contribution to the direct heating with a dotted line at a ratio of
unity. In the bottom panel shows the $X_{56}$ distribution. In the
bottom right plot we consider a shallower distribution of $^{56}$Ni,
as can be seen its the bottom panel.

These example confirm our general conclusion that when the $^{56}$Ni
is more shallow or has a shallower slope, the impact of the
diffusive tail is diminished. These examples also show that in the
shallow outer layers exposed at early times, the diffusive tail is
always at least comparable to the direct $^{56}$Ni luminosity, so
that $L_{56}$ cannot be inferred from the observations more
accurately than a factor of $\sim2$. This uncertainty is not overly
problematic because of the number of other estimates in this work
and estimates of $M_{56}$ within an order of magnitude will still
facilitate our goals. Such a limitation should not be surprising,
because the diffusive tail always leaks at least a little and
contaminates $L_{56}$. Our investigation of this effect also shows
that the ratio $L_{\rm tail,total}/L_{56}$ strongly depends on the
prefactor in equation (\ref{eq:tailtotal2}), which comes from the
power-law scaling of $\Delta M_{\rm diff}\propto t^\eta$. We
generally find that this ratio is weaker with smaller $\eta$,
demonstrating that $L_{56}$ can be estimated more reliably in such
circumstances.

\section{The Diffusion Depth}
\label{sec:diffusiondepth}

Here we derive the depth of the diffusion wave as a function of
time, as was used in equation (\ref{eq:mdiff}). This roughly matches
the derivation by \citet{RabinakWaxman2011}, with a few changes of
numerical factors.

The pre-explosion density profile of the star is approximated as \be
    \rho_0(r_0) = \rho_{1/2}\delta^n,
\ee where $\rho_{1/2}$ is the density at half the radius,
$\delta=1-r_0/R_*$ is the fractional radius, and $n$ is the
polytropic index. The fraction of ejecta mass lying above a radius
$r_0$ is \be
    \delta_m = \frac{1}{M}\int_{(1-\delta)R_*}^{R_*} 4\pi r_0^2 \rho_0(r_0)dr_0 \approx \frac{3f_\rho}{n+1}\delta^{n+1},
\ee so that the mass depth is $\Delta M=\delta_mM$.

The velocity profile from the passage of the SN shock is taken from
\citet{MatznerMcKee1999}, who use an interpolation between the
Sedov--von Neuman--Taylor and the
Gandel'Man--Frank-Kamenetskii--Sakurii self-similar solutions
\citep{VonNeumann1947,Sedov1959,Taylor1950,GandelManFrankKamenetskii1956,Sakurai1960}.
Near the surface of the star this gives \be
    v_s = A_v \lp \frac{E}{M} \rp^{1/2} \lp\frac{4\pi}{3f_\rho} \rp^\beta \delta ^{-\beta n},
\ee where $A_v\approx0.8$, $\beta=0.19$, and
$f_\rho=\rho_{1/2}/\rho_*$ is the ratio of the half-radius density
to the average density $\rho_*=3M/4\pi R_*^3$. In Appendix
\ref{sec:structurefactor} we derive $f_\rho$ for some appropriate
density profiles for the SN I progenitors we are interested in.
\citet{MatznerMcKee1999} show that the final velocity of the fluid
is given by $v_f (r_0)= f_v(r_0)v_s(r_0)$, where $f_v\approx2$.

The thermal diffusion length scale is \be
    D(\delta_m,t) \approx \sqrt{\frac{ct}{3\kappa \rho(\delta_m,t)}},
\ee where the density is given from continuity to be \be
    \rho = - \frac{M}{4\pi r^2 t} \lp \frac{d v_f}{d\delta_m} \rp^{-1}\approx \frac{1+1/n}{\beta} \frac{\Delta M}{4\pi t^3 v_f^3}.
\ee The diffusion depth is approximated as the location where $D$ is
roughly equal to the thickness of a given shell of material \be
    \Delta r = -t\frac{d v_f}{d\delta_m}\approx \frac{\beta}{1+1/n}\frac{v_f}{\delta_m}.
\ee Setting $\Delta r=D$, we find \be
    \delta_m = \lb \frac{1+1/n}{\beta} \frac{f_vA_vc}{3\kappa} \frac{4\pi t^2}{M} \lp\frac{E}{M}\rp^{1/2}
        \lp\frac{4\pi}{3f_\rho}\rp^\beta \lp\frac{3f_\rho}{n+1}\rp^{\beta/(1+1/n)} \rb^{(1+1/n)/(1+1/n+\beta)}.
\ee This expression is then used to derive equation
(\ref{eq:mdiff}). Note that $\Delta M_{\rm diff} = M\delta_m\propto
f_\rho^{-0.04}$ for $\beta=0.19$ and $n=3$. In Appendix \ref{sec:structurefactor} we
find that $f_\rho$ ranges from $\approx0.01$ for SN Ib/c progenitors
(this is what is used for the numerical factor in eq.
[\ref{eq:mdiff}]) to $\approx 1$ for SN Ia progenitors, so the
prefactor of equation (\ref{eq:mdiff}) varies by $\approx20\%$
between these two very different scenarios.

\section{The Density Structure Factor}
\label{sec:structurefactor}

Even though the diffusion depth is relatively insensitive to the
dimensionless density structure factor $f_\rho$, it is worth
approximating just how greatly it varies for different SN I
progenitors. Using \citet{MatznerMcKee1999}, for a radiative
envelope of constant opacity \be
    f_\rho = \frac{\pi}{144} \frac{\beta^4}{1-\beta} \lp\frac{\mu m_p}{k_{\rm B}}\rp^4 G^3M^2a
\ee where $1-\beta=L_*/L_{\rm Edd}$ is the ratio of the stellar
luminosity to the Eddington limit. For a Thomson opacity with
hydrogen-deficit material this ratio is \be
    1-\beta = 0.51 \lp\frac{L}{10^5L_\odot} \rp \lp\frac{M}{3M_\odot}\rp^{-1}.
\ee This then gives $f_\rho\approx0.015$ for a SN Ib/c progenitor.

For degenerate electrons as in a WD progenitor for SNe Ia, the
equation of state is $P=K\rho^{1+1/n}$. For the case of
non-relativistic electrons $n=3/2$ and
$K=9.91\times10^{12}\mu_e^{-5/3}$, and for relativistic electrons
$n=3$ and $K=1.23\times10^{15}\mu_e^{-4/3}$, where $\mu_e$ is the
molecular weight per electron and $K$ is in cgs units. In this case,
\be
    f_\rho = \frac{4\pi R_*^3}{3M} \lb\frac{GM}{(n+1)KR_*} \rb^n \approx 1.3 \lp\frac{M}{1.4M_\odot} \rp^2,
\ee where in the last expression we have assumed relativistic
electrons and $\mu_e=2$.

\section{The Thermalization Depth}
\label{sec:Thermalizationdepth}

Here we approximate the optical depth at the thermalization depth as
a function of the observed color temperature, $T_c$, the
photospheric velocity and the time since explosion. The
thermalization depth is defined as the outermost region which is in
thermal equilibrium. In case that scattering dominates the opacity,
and assuming that free-free dominates the photon generation (and
opacity), this is the location where free-free emission generates
within a diffusion time just enough photons to sustain thermal
equilibrium (see \citealt{NakarSari2010} for a detailed discussion).
Namely, $n_{\rm BB}(T_c)=t_{\rm diff} \dot{n}_{\rm ph,ff}(T_c)$ where
$n_{\rm BB}(T_c) \approx aT_c^4/3k_{\rm B}T_c$ is the blackbody photon density
and $\dot{n}_{\rm ph,ff} \approx 3 \times 10^{36} {\rm~s^{-1}~cm^{-3}}
\rho^2 T_c^{-1/2} x_i$ is the volumetric rate at which photons with
temperature $\sim T_c$ are generated by free-free (in cgs units), $a$
is the radiation constant, $k_{\rm B}$ is Boltzmann's constant, $\rho$ is
the mass density at the thermalization depth and $x_i$ is the
average number of ionized electrons per atom. Now, the diffusion
time is $t_{\rm diff} \approx (\tau_c r_c/c) \beta/(1+1/n)$
while $\tau_c = \kappa \rho r_c \beta/(1+1/n)$. Therefore, taking
$\beta=0.19$ and $n=3$ and neglecting the difference between $v_c$
and $v_{\rm ph}$, which introduces completely negligible correction, we
obtain:
\begin{equation}\label{eq:tauc}
    \tau_c \approx 2 \left(\frac{T_c}{10^4
    \rm{~K}}\right)^{7/6} \left(\frac{v_{\rm ph}}{10^4
    {\rm ~km/s}}\right)^{1/3} \left(\frac{t}{\rm{day}}\right)^{1/3},
\end{equation}
where we approximate $x_i^{-1/3}=1$ and $\kappa_{0.1}^{2/3}=1$ ,
which are reasonable values for the typical early temperatures of
$T_c=10,000-20,000$ K. For lower $T_c$ both $x_i$ and $\kappa$,
(which are linear at each other) are  smaller and so is $\tau_c$.
This estimate of $\tau_c$ assumes that bound-free absorption is
negligible, otherwise $\tau_c$ is smaller. Finally, $\tau_c>1$ is
always satisfied. If equation (\ref{eq:tauc}) indicates otherwise, then
its assumptions are not valid (e.g., scattering does not dominate
the opacity) and $\tau_c \approx 1$.


\end{appendix}

\end{document}